\documentclass[12pt,showpacs,preprintnumbers,floatfix]{article}
\pdfoutput=1
\usepackage{graphicx} 
\usepackage{dcolumn} 
\usepackage{bm} 
\usepackage{amsfonts}
\usepackage{amsthm}
\usepackage{amsmath}
\usepackage{amssymb}
\usepackage{epsfig}
\usepackage{color}
\usepackage{textcomp}
\usepackage{hyperref}
\usepackage{subfigure}
\usepackage[utf8]{inputenc}
\usepackage[english]{babel}
\usepackage[usenames,dvipsnames,table]{xcolor}
\usepackage{enumitem}
\usepackage{mathabx}

\def\be{\begin{equation}}
\def\ee{\end{equation}}
\def\ba{\begin{eqnarray}}
\def\ea{\end{eqnarray}}

\def\bdm{\begin{displaymath}}
\def\edm{\end{displaymath}}

\def\bq{\begin{quote}}
\def\eq{\end{quote}}

 at 10truept
\setlist[itemize]{leftmargin=2cm,rightmargin=1cm}

\newcommand{\bea}{\begin{eqnarray}}
\newcommand{\eea}{\end{eqnarray}}

\newcommand{\bi}{\begin{itemize}}
\newcommand{\ei}{\end{itemize}}

\newcommand{\beq}{\begin{equation}}
\newcommand{\eeq}{\end{equation}}
\newcommand{\beqa}{\begin{eqnarray}}
\newcommand{\eeqa}{\end{eqnarray}}

\newcommand{\mU}{\mathcal{U}}
\newcommand{\tot}{\text{tot}}

\def\ltap{\ \raise.3ex\hbox{$<$\kern-.75em\lower1ex\hbox{$\sim$}}\ }
\def\gtap{\ \raise.3ex\hbox{$>$\kern-.75em\lower1ex\hbox{$\sim$}}\ }
\def\gl{\ \raise.5ex\hbox{$>$}\kern-.8em\lower.5ex\hbox{$<$}\ }
\def\roughly#1{\raise.3ex\hbox{$#1$\kern-.75em\lower1ex\hbox{$\sim$}}}

\begin{document}

\thispagestyle{empty}
\begin{flushright}
May 2025
\end{flushright}
\vspace*{1.7cm}
\begin{center}
{\Large \bf Multiverse Predictions for Habitability:\\ The Habitability of Exotic Environments}\\

\vspace*{1.2cm} {\large McCullen Sandora\footnote{\tt
mccullen@bmsis.org}}\\
\vspace{.5cm} 
{\it  $^{}$Blue Marble Space Institute of Science}\\
	{\it  Seattle, WA 98154, USA}

\vspace{2cm} ABSTRACT
\end{center}
The relative abundances of exotic environments provides us with (uninformed) bounds on the habitability of those environments relative to our own, on the basis that our presence here is not too atypical.  For instance, since red stars outnumber yellow stars 7 to 3, we can infer that red stars must be less than 8.1 times as habitable as yellow, as otherwise our presence around a yellow star would be a statistical outlier at the level of $5\%$.  In the multiverse context, the relative abundances of exotic environments can be drastically different from those in our universe, which sometimes allows us to place much stronger bounds on their relative habitability than we would get by restricting our attention to our universe.  We apply this reasoning to a variety of different exotic environments: tidally locked planets, binary star systems, icy moons, rogue planets, liquids with properties different from water, and waterworlds.  We find that the bounds on the relative habitability of rogue planets and waterworlds are at least an order of magnitude stronger in a multiverse context than from our universe alone. Additionally, the belief that some of water's special properties are essential for life, such as the fact that ice floats and, with some caveats, that it acts as a universal solvent, are incompatible with the multiverse hypothesis.  If any of these bounds are found to be violated in the future, the multiverse hypothesis can be falsified to a high degree of confidence.

\vfill \setcounter{page}{0} \setcounter{footnote}{0}
\newpage


\section{Introduction}

This is a continuation of work aimed at using the multiverse hypothesis to generate concrete, testable predictions for the habitability of different environments throughout the universe.  Up to this point, these predictions were created primarily by focusing on the probabilities of observing our particular physical constants within a multiverse context, which can be quite sensitive to assumptions made about habitability conditions. Demanding that the probability of our observations is not too atypical gives us predictions for which habitability conditions are correct, which can then be checked through exploration of our universe.  This formalism has been applied to a variety of domains to successfully generate predictions, including stellar properties \cite{mc1,mc7}, planetary properties \cite{mc2,mc6}, element abundances \cite{mc5}, and life's origin \cite{mc8} and evolution \cite{mc3,mc4}.

Here, we extend this analysis to not only consider the probability of observing our constants, but also the probability that we are humanlike, in the sense that we observe our environment to possess its particular characteristics, as opposed to several other types of exotic environment. In general, this probability is dependent on both the abundance and relative habitability of each of these environments.  Since the relative abundances of the various exotic environments we consider can be determined both in our universe and within the multiverse context, demanding that our observation as being humanlike is typical allows us to place constraints on the relative habitability of these exotic environments.

This type of reasoning is already being employed when discussing the habitability of various types of locations within our universe.  One prominent example is the discussion of whether life can possibly evolve around red dwarf stars, which outnumber yellow stars by a factor of $70/30$.  If life is possible around red stars, that would make our orbit around a yellow star atypical \cite{yellowinstead}, though not unreasonably so; our location around a yellow star is within the 95$\%$ confidence interval as long as the relative habitability of red dwarf stars is less than 19$\times$3/7=8.1 that of yellow stars.  While this is not very constraining, the assumptions one makes about habitability can alter these conclusions substantially. For instance, as pointed out in \cite{yellowinstead}, if the probability of life arising on a planet is proportional to the stellar lifetime, the relative habitability must be much lower for red dwarf stars in order to explain our presence around a yellow star.  If instead Carter's hard step model \cite{carterbio} is adopted, where the probability of emergence of complex, intelligent life is proportional to the stellar lifetime raised to some integer power, our occurrence around a yellow star is indeed highly atypical \cite{waltham2017star}.  In either case, the longevity of red stars implies that the majority of instances of life would occur at a much later time in the evolution of the universe, making our temporal position unlikely \cite{slbtime}.

This analysis can be repeated for any type of exotic environment.  In each case, recent results in astronomy and planetary science allow us to estimate the relative number of these locations relative to Earthlike planets.  However, in many of these cases the habitability of each of these systems is poorly constrained.  Our results then offer theoretical predictions for the habitability of these environments, to ensure that our existence at this particular type of location is typical.

Coupling this analysis with the multiverse framework can sometimes allow us to place stronger bounds on the relative habitability of certain environments than when attention is restricted to our universe.  This is because the relative abundances within the multiverse context can be greatly different than within our universe.  The condition for a more stringent bound comes about when the relative proportion of these environments is much larger in the multiverse context than in our universe.  Indeed, for some of the exotic environments we consider, no strong bounds are placed on their relative habitability based on our universe alone, but strong bounds are placed in the multiverse context.  This occurs when these environments do not greatly outnumber Earthlike conditions in our universe, but when considering other potential values of the fundamental constants these exotic environments become much more prevalent.

Stated another way, this allows us to make predictions for which type of environments will be relatively uninhabitable, if we take the multiverse hypothesis to be true: those which dominate in other types of universes.  This is interesting precisely because though unknown at the moment, the habitability of many of these alternative environments will be determined eventually, some of which in the relatively near future.

We apply this logic to several different exotic environments in turn, placing some bounds on the relative habitability of each of them, if the multiverse hypothesis is taken to be true. We start with tidally locked planets in section \ref{TL} as a simple scenario in which we have worked out the details previously.  We then use this formalism to place bounds on the relative habitability of binary star systems in section \ref{binary}, the relative habitability of icy moons versus rocky planets in section \ref{icy}, of rogue planets in \ref{rogue}, and the habitability of liquids other than water, and waterworlds versus rocky planets in section \ref{water}.  The bounds on relative habitability are summarized in Table \ref{r_table} in the Discussion section.

Continuing with our general program, these predictions will serve as guideposts that can eventually be compared with information we will be able to glean about the habitability of these environments in the future, and so offer a slew of potentially testable predictions of the multiverse hypothesis.

\subsection*{Probabilities}

\begin{figure*}
\begin{center}
\includegraphics[width=1.0\textwidth]{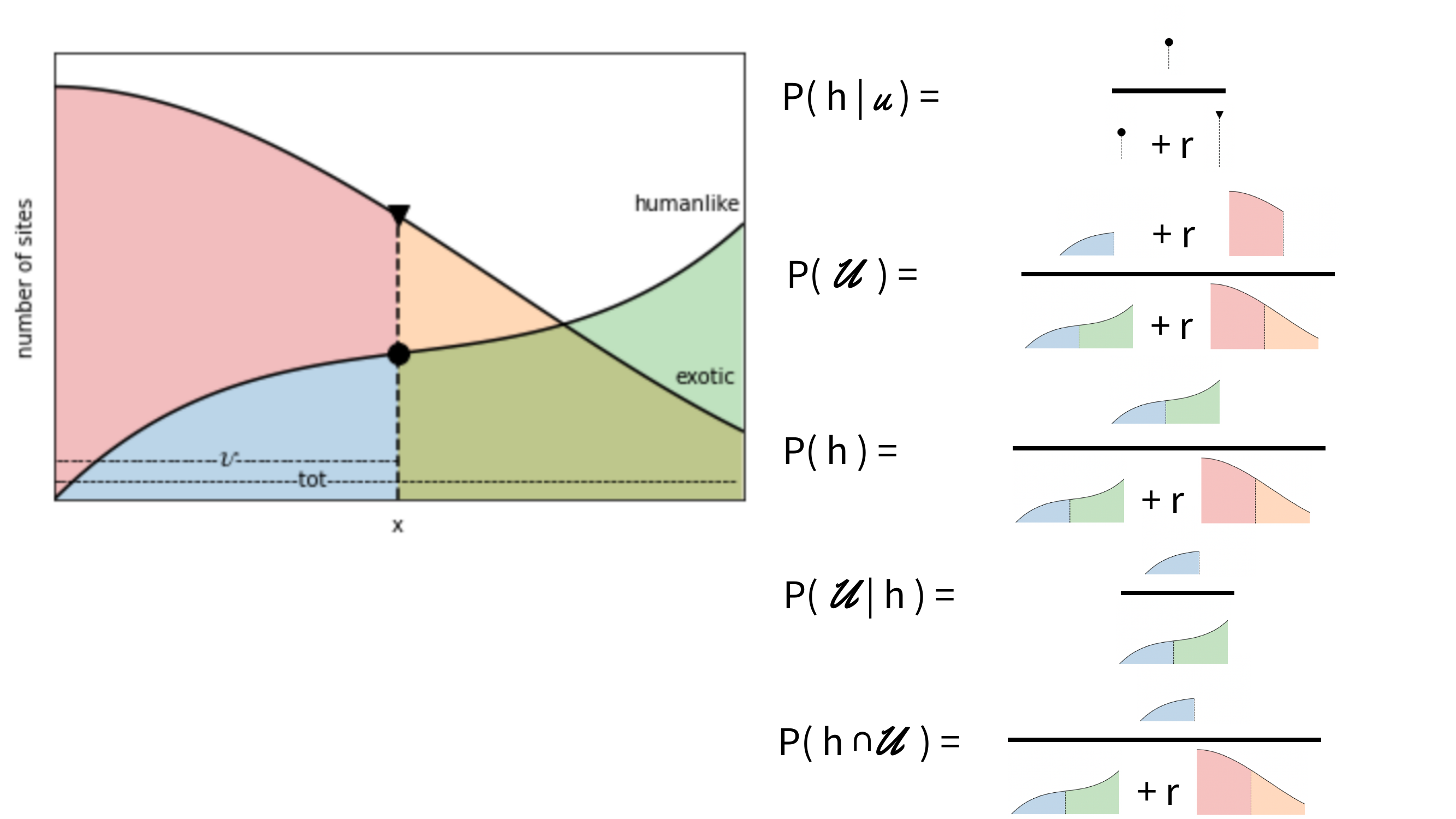}
\caption{Illustration of the various probabilities discussed. The variable $x$ may refer to any of the physical constants $\alpha$, $\beta$, $\gamma$ etc., with the others integrated over.  The dashed lines indicate our value of $x$.  The region $\mU$ designates universes like ours, in that the constant $x$ is no greater than what we observe.}
\label{shapefig}
\end{center}
\end{figure*}

Our previous works mostly calculated the probability of being in our universe, given a habitability condition $\mathbb H$.  Here, we augment this original computation with the probability that we are humanlike as well.  Let us consider the simple case where there are observers in two qualitatively different kinds of environments, dubbed humanlike and exotic observers.  The number of sites for each are denoted $S_h$ and $S_e$, respectively, and the number of observers of each is related to habitability by definition $N_h=H_hS_h$, $N_e=H_eS_e$.  The relative habitability is defined as $r=H_e/H_h$.  It will often be relatively simple to compute $S_h$ and $S_e$ from physical principles, but the relative habitability factor is unknown.  Therefore, we can place confidence limits on how habitable this exotic type of environment is, given the fact that we are humanlike.  The different probabilities discussed in this section are illustrated in Fig. \ref{shapefig}.

Let us first restrict our attention to observers within our universe.  This would be a relevant quantity irrespective of the multiverse hypothesis.  If we denote quantities within our universe with the superscript $u$ and the quality of being humanlike as $h$, then the probability of being humanlike when restricting attention to our universe is
\beq
P(h\,|\,u)=\frac{S^u_h}{S^u_h+r\,S^u_e}
\eeq
If we demand our observations (of being humanlike) to be statistically typical to some degree $p_c$ (which we generally take to be .05), this places an upper limit on the relative habitability:
\beq
r^{h|u}<\left(\frac{1}{p_c}-1\right)\frac{S^u_h}{S^u_e}\label{rhu}
\eeq
These limits are strongest when the number of exotic sites is much larger than the number of humanlike sites.

The multiverse hypothesis adds additional restrictions to the relative habitability, because the ratio of exotic to humanlike sites in the multiverse setting may be significantly different from that of our universe.  For this, we extend our analysis by considering an ensemble of universes.  Here we denote the relevant factors in each universe by $x$, which in general is a set of continuous variables corresponding to the fundamental constants specifying the universe under consideration.  The constants we vary are the fine structure constant $\alpha$, the ratio of the electron to proton mass $\beta=m_e/m_p$, the ratio of the proton mass to the Planck mass $\gamma=m_p/M_{pl}$, and the ratio of the up and down quark masses to the proton mass, $\delta_u=m_u/m_p$ and $\delta_d=m_d/m_p$.  We also make use of the parameter $\kappa=1.1\times10^{-16}$ that characterizes the galactic density as $\rho_\text{gal}=5.6\times10^7\kappa^3m_p^4$, though in this paper we hold it fixed.  Lastly, we often refer to the stellar mass of a system, denoted here as a dimensionless ratio to the Chandrasekhar mass $\lambda=M_\star/((8\pi)^{3/2}M_{pl}^3/m_p^2)$. 

In the multiverse context we may also ask: What is the probability that we are in a universe like this one? The concept of ``like this one" is ambiguous- here it is defined as a condition $\mU$ on the physical constants representing the probability that we observe a value of a particular constant to be at least as large as ours.  So, we will have a different condition $\mU$ for each of the five physical constants we vary in this paper, and generically take the most constraining condition for each case.  Then we have
\beq
P(\mathcal U)=\frac{\int_\mU dx\, S_h(x)+r\,S_e(x)}{\int_\tot dx\, S_h(x)+r\,S_e(x)}\equiv\frac{S^\mU_h+r\,S^\mU_e}{S^\tot_h+r\,S^\tot_e}\label{PU}
\eeq
Here, we have denoted with superscripts $\mU$ and $\tot$ the total number of sites satisfying condition $\mU$ and throughout the multiverse, respectively.  

Above, we have made the assumption that the relative habitability of the two types of sites is independent of the physical constants.  This resultant probability is bounded between $S^\mU_h/S^\tot_h$ and $S^\mU_e/S^\tot_e$, so if both of these are above (or below) the desired confidence limit then nothing can be said about the relative habitability $r$.  Otherwise if only one quantity is below this value this places limits on $r$.
\beq
r^\mU<\frac{S^\tot_h}{S^\tot_e}\,\frac{\frac{S^\mU_h}{S^\tot_h}-p_c}{p_c-\frac{S^\mU_e}{S^\tot_e}}
\eeq

Considering only this set of probabilities, and comparing the two cases $r=0$ and $r=1$,  has already allowed us to place some stringent limits on possible habitability conditions in our previous papers.  However, this procedure can be augmented by considering some additional probabilities as well.  The first is the probability that we are humanlike within the multiverse context:
\beq
P(h)=\frac{\int_\tot dx\, S_h(x)}{\int_\tot dx\, S_h(x)+r\,S_e(x)}\equiv\frac{S^\tot_h}{S^\tot_h+r\,S^\tot_e}
\eeq
giving
\beq
r^h<\left(\frac{1}{p_c}-1\right)\frac{S^\tot_h}{S^\tot_e}\label{rU}
\eeq
The conditional probability that we are in a universe such as this one (satisfying condition $\mU$) given that we are human is
\beq
P(\mathcal U\,|\,h)=\frac{\int_\mU dx\, S_h(x)}{\int_\tot dx\, S_h(x)}\equiv\frac{S^\mU_h}{S^\tot_h}
\eeq
This is equivalent to the probability of being in our universe, equation \ref{PU} under the most stringent habitability assumption that only humanlike environments are habitable, with $r=0$.

Finally, we may consider the joint probability that we are human and within this universe as
\beq
P(h\cap\mathcal U)=\frac{\int_\mU dx\, S_h(x)}{\int_\tot dx\, S_h(x)+r\,S_e(x)}\equiv\frac{S^\mU_h}{S^\tot_h+r\,S^\tot_e}
\eeq
This is, of course, equal to the product of the previous two probabilities, and so will always be smaller than and thus more constraining than either.  It is also strictly smaller than $P(\mU)$, and so, provided $\frac{S^\mU_h}{S^\tot_h}>p_c>\frac{S^\mU_e}{S^\tot_e}$, will place the strongest bounds on $r$ between these two conditions.

Demanding that this probability exceeds the threshold $p_c$ leads to the following condition on the relative habitability of exotic environments:
\beq
r^{h\cap\,\mU}<\left(\frac{1}{p_c}-1\right)\frac{S^\mU_h}{S^\tot_e}
\eeq

\section{Tidal Locking}\label{TL}

As a simple first example, let us consider the relative habitability of tidally locked planets, both within our universe and within the multiverse context.  The details of how tidal locking depends on physical constants were worked out in \cite{mc1}, leaving us free to use these results to illustrate the approach we take.

Tidally locked planets orbit their host star in such a way that one side is permanently facing the star.  This occurs due to rotational spindown, which results from tidal forces from the host star over the lifetime of the planetary system; tidal locking of temperate planets thus occurs preferentially around lower mass stars, where planets orbit closer to their star and tidal forces are stronger.  In \cite{mc1} it was found that there is a stellar mass threshold, below which temperate, terrestrial planets are expected to undergo tidal locking, which is
\beq
\lambda_\text{TL}=.89\,\alpha^{5/2}\,\beta^{1/2}\,\gamma^{-4/11}
\eeq
This is normalized to $.85M_\Sun$ for our values of the constants.

It has traditionally been argued that tidally locked planets are not habitable, as temperature extremes between the day and night sides preclude an appreciable temperate region from developing \cite{barnesTL}.  However, optimists have considered the habitability of these worlds \cite{1997Icar..129..450J}.  We may augment the modeling work being done to address this issue by a purely probabilistic argument taking into account only the relative abundances of tidally locked versus tidally unlocked worlds.

If we only consider bare numbers, then, factoring in the distribution of stellar masses, within our universe about $86\%$ of stars are below the tidal locking threshold, so that their relative abundance is about 6 to 1.  If we also take into consideration the propensity of each stellar mass to form planets along the lines of \cite{mc2}, we find that in order for the probability of finding ourselves on a tidally unlocked planet to exceed $p_c=.05$, the relative habitability of tidally locked planets must be $r^{h|u}_\text{TL}<3.9$ from eqn (\ref{rhu}).  When considering the relative abundances within a multiverse context, the probability of being on a tidally unlocked world exceeds $p_c=.05$ when $r^h_\text{TL}<23.5$.  However, with these assumptions, the probability of being a universe with gravitational strength $\gamma$ as small is ours is $P(\gamma\leq\gamma_\text{obs})=4.5\times10^{-7}$, and so the probability of being both in a universe like ours and being on a tidally unlocked world never exceeds $p_c=.05$, no matter the relative habitability \footnote{All calculations performed in this paper use the code from \url{https://github.com/mccsandora/Multiverse-Habitability-Handler}}.

We've encountered this problem before in \cite{mc1}, where it was found to stem from the assumption that all stars are equally habitable, regardless of their properties.  There and in \cite{mc5}, we found that if we instead use the habitability conditions that complex life can only arise around yellow stars (denoted by boldface {\bf yellow}), in universes with carbon to oxygen ratio C/O$<1$ (denoted {\bf C/O}), and that the habitability of a planet is directly proportional to the total amount of entropy it processes over its lifetime (denoted {\bf S}), the probabilities of observing our values of the constants $P(\mU)$ all exceed .17, making our observations consistent with the multiverse hypothesis.  We can augment these considerations with the probability of being on tidally unlocked planets.  With the strict yellow starlight consideration, when restricting to our universe, we find the very weak bound $r^{h|u}_\text{TL}<507.4$, as the vast majority of stars that induce tidal locking are red dwarfs, and so are at the outset considered to be uninhabitable anyway.  In the multiverse context, we have $r^h_\text{TL}<2.0$, and $r^{h\cap\,\mU}_\text{TL}<.38$, so the bounds on relative habitability are orders of magnitude stronger than when considering our universe in isolation.

Because this treatment explicitly excludes most sites where tidal locking takes place in our universe, we consider how the bounds are altered when we relax the condition that the host starlight be yellow, and instead enforce a more encompassing photosynthetic condition from \cite{mc1} which includes red dwarfs.  In this case, we find $r^{h|u}_\text{TL}<.91$, $r^h_\text{TL}<.68$, and $r^{h\cap\,\mU}_\text{TL}<.13$.  So, even though the bounds from our universe are greatly affected, the bounds from the multiverse are less sensitive, and are still shown to provide stronger constraints.  These conclusions are similarly robust if we alter the {\bf C/O} condition.

This illustrates the general procedure for placing bounds on the relative habitability of any environment, both from within our universe and within the multiverse framework.  Attention should be restricted to habitability assumptions that are known to be compatible with our presence in this universe, and so, unless otherwise specified, we favor the {\bf yellow + C/O + S} habitability condition.  We now extend this analysis to other types of exotic environments.

\section{Binary Systems}\label{binary}

We now turn our attention to binary stellar systems.  In all our previous works, these were implicitly ignored, but in fact they comprise a significant fraction of all stellar systems.  Though the binary fraction depends on stellar mass, the overall percentage of stars in binary systems is close to $50\%$ \cite{duchene2013stellar}.  This fraction depends on fundamental constants, and as such the overall habitability of a universe can depend on the relative habitability of binary systems.

There has been much discussion on the habitability of binary systems.  Arguments have been made that planets in these systems may be much less likely to be habitable, on the basis that secular perturbations from the companion star would destabilize orbital parameters \cite{kaib2013planetary}, an effect that would particularly afflict planets orbiting both stars (on circumbinary orbits) \cite{simonetti2020statistical}.  On the other hand, it was argued in \cite{holman1999long} that the presence of a companion star would not greatly affect the orbital stability of temperate zone planets, as long as the binary semimajor axis was either $<.3$AU or $>10$AU.  Additionally, it was argued in \cite{shevchenko2017habitability} that binary systems may in fact be superhabitable, as a variety of desired habitability properties that only arise in special circumstances for planets orbiting single stars automatically hold for planets orbiting binaries.  Lastly, even if the habitability of binary systems does not differ from single stars, a higher binary fraction may significantly impact overall habitability, due to the fact that the gravitational cross section of binary systems is significantly larger, leading to enhanced planetary orbit destabilization of neighboring stars \cite{kaib2014very,li2015cross}.  In this work we remain agnostic, instead treating the relative habitability of binary systems as a free parameter $r_\text{binary}$, which we proceed to place limits on based on the binary star fraction, both within our universe and within the multiverse.

Most stars form within stellar clusters, and in this context most stars will be gravitationally bound to at least one other star at inception \cite{korntreff2012towards}.  Therefore, when computing the binary stellar fraction, we are trying to determine the fraction of systems that remain undisrupted for the duration of their evolution.  If a disruption event is to occur, it will most likely be within the initial birth cluster, as evidenced by the steady decline of binary fraction for systems with very young stellar age \cite{li2020encounters}, and near constant fraction thereafter \cite{duchene2013stellar}.  These disruptions also significantly alter the distribution of binary separations; this begins as a log-uniform distribution (reflecting equipartition) and then evolves into a peaked distribution by gas dynamics and cluster dynamics eroding binary systems with short and long periods, respectively \cite{korntreff2012towards}.  The resulting typical separation distance $a_\text{bin}$ was calculated in 
\cite{low1976minimum}, where it was found to be set by the Jeans length of the cluster.  In that same paper they suggest the distribution of separations to be a Rayleigh distribution.  Though others have suggested instead a log-normal distribution of separations \cite{korntreff2012towards}, in this work we utilize the Rayleigh for its analytic properties, and in the absence of any clear theoretical motivation otherwise.

The fraction of systems that remain in binaries will then depend on the optical depth of the birth cluster $\tau=\xi\pi nvta^2$.  Here $t$ is the typical cluster lifetime, $v$ typical cluster speed, $n$ cluster density, $a$ the binary separation, and $\xi\sim\mathcal O(1)$ is a constant encapsulating the details of the interaction and cluster properties.  This fraction can be computed as
\beq
f_\text{bin}=\int_o^\infty da \frac{a}{a_\text{bin}^2}\,e^{-a^2/(2a_\text{bin}^2)-\xi\pi nvta^2}
\eeq
 Which evaluates to 
\beq
f_\text{bin}=\frac{1}{1+2\pi\,\xi\,n\,v\,t\,a_\text{bin}^2}
\eeq
Using $a_\text{bin}\sim (M_\text{Jeans}/\rho_\text{gal})^{1/3}$, $n\sim\rho_\text{gal}/M_\star$, $v\sim G^{1/2}M_\star^{1/3}\rho_\text{gal}^{1/6}$, and $t\sim(G\rho_\text{gal})^{-1/2}$, we find $a_\text{bin}^2nvt\sim (M_\text{Jeans}/M_\star)^{2/3}$.  Using the expression for the Jeans mass $M_\text{Jeans}\sim m_e^{2/7}M_{pl}^{20/7}/(\alpha^2m_p^{15/7})$ from \cite{low1976minimum} and fixing the value of $\xi$ to reproduce the binary fraction observed in our universe, we arrive at\footnote{As a disclaimer, if we instead use the parameterization of the typical binary separation from \cite{heggie1975binary}, where it is stated this is dictated by the typical cluster energy, $GM/a_\text{bin}\sim T/m_p$, we find that the binary fraction is independent of constants. If this scenario is true, the binary fraction is not set by competing physical processes, but instead robustly tends to be relatively equal to the lone star population.  In this case, the reader may disregard the conclusions of this section.}
\beq
f_\text{bin}=\frac{1}{1+29.1\,\alpha^{-25/21}\,\beta^{29/42}\,\gamma^{2/21}}
\eeq

With these, the relative habitability of binary star systems, when restricted to our universe, is
\beq
r^{h|u}_\text{binary}<19.0
\eeq
When we consider throughout the multiverse, we instead have
\beq
r^h_\text{binary}<25.9,\quad r^{h\cap\,\mU}_\text{binary}<5.0
\eeq
So that the bounds on $r$ are somewhat lessened just considering the probability of being around a lone star within the multiverse context, reflecting the fact that binary star systems are a more common in our universe compared to the average throughout the multiverse.  However, when considering the joint probability of being both human and in this universe, the bounds on the relative habitability of binary systems becomes nearly 4 times stronger than they are from just considering the relative populations within our universe alone.
	
\section{Icy Moons}\label{icy}

Another potentially significant contribution to the overall habitability of a universe may come from the icy moons orbiting gas giant planets, many of which possess subsurface oceans beneath a thick ice crust.  There is reason to be skeptical that these worlds may be capable of supporting complex life.  For one, the thick ice covering prevents photosynthesis, which limits the size of any potential biosphere to less than 1,000 times smaller than it otherwise would have been able to attain \cite{des2000did}. Underwater hydrothermal vents may provide a source of energy, but the total biomass from these would again be diminished by a factor of at least $10^8$ \cite{mccollom1999methanogenesis}.  Subsurface oceans would also have an extremely limited oxygen content, which on Earth is necessary for the most productive metabolisms, such as for animals \cite{gaidos1999life}.

However, there is also reason to be optimistic about the prospects of these worlds.  While an ice cover may stymie photosynthetic activity, cracks in the ice are inevitable, and are potential locations for photosynthesis \cite{chyba2002europa}.  As was pointed out in \cite{mojzsis2022geoastronomy}, within the solar system, the total volume of liquid water is dominated by subsurface oceans in the outer system.  And, while Jupiter-like planets may only exist around $27\%$ of stars, they host multiple moons, providing potentially even more opportunities for life to emerge than on rocky interior planets \cite{fernandes2019hints}.  \footnote{Another estimate gives the cold gas giant occurrence rate as $6.7\%$, though the disparity is due more to the differences in definition, rather than methods. \cite{wittenmyer2020cool}.  This can be compared to the Earthlike planet occurrence rate, which was estimated as $37\%$, with average number .6 Earthlike planets per star in \cite{bryson2020occurrence}.  Given the ambiguities in the definitions of these values, the uncertainties in their measurement, and that they are actually quite similar, we neglect these factors when estimating the relative abundance of Jupiterlike and Earthlike planets.}

In order to assess the abundance of icy moons as a function of fundamental constants, we estimate the typical number of gas giants per stellar system, and multiply this by the typical number of moons orbiting gas giants.  Additionally, we find a condition for a moon to have a tidally induced subsurface ocean based on tidally induced heating.

For this section and the next, we briefly describe the standard planet formation setup, described in more detail in the multiverse setting in \cite{mc2,mc6}.  A star begins with a circumstellar disk with surface density $\Sigma(a)=\xi_\text{disk}M_\star/(2\pi a_\text{disk}a)$, where $a$ is the semi-major axis, $a_\text{disk}\sim (M_\star/\rho_\text{gal})^{1/3}=3.6\times10^{-6}\lambda^{1/3}M_{pl}/(\kappa m_p^2)$, and $\xi_\text{disk}\sim.01$ parameterizes the mass of the disk relative to the star.  This is in accordance with the initial distribution of material present in protoplanetary disks, and matches the resulting distribution of observed planetary orbits \cite{williams2011protoplanetary}.  In the initial stages of planet formation, planets form by accreting gas, dust, pebbles and planetesimals within their basin of attraction $\Delta a$, yielding bodies of mass $M\sim2\pi a \Delta a\Sigma$.  If $\Delta a$ is taken to be proportional to the planet's Hill radius, which defines its zone of gravitational dominance, $R_\text{Hill}=(M_\text{planet}/(3M_\star))^{1/3}a$, this yields the isolation mass, $M_\text{iso}\sim M_\star (a\xi_\text{disk}/a_\text{disk})^{3/2}$.  This arrangement is ultimately unstable, yielding to a phase of giant impacts, where the basin of attraction is given by $\Delta a \sim 2 v_\text{esc}/\Omega$, with $v_\text{esc}$ the escape velocity and $\Omega\sim \sqrt{Gm_\star/a^3}$ the Kepler frequency.  This yields the oligarchic mass $M_\text{olig}\sim5.4\times10^6\kappa^{3/2}\lambda^{1/4}\alpha^{3/4}m_e^{3/4}m_p^{7/4}M_{pl}^{3/4}a^{9/4}$, which sets the typical mass scale of the resulting planets.

In \cite{mc6}, following \cite{ginzburg2019end}, we found that the typical gas giant mass was given by $M_\Jupiter\sim2.4\times10^4\kappa^{1/3}\lambda^{2.01}M_{pl}^{2.43}/(\alpha^{4.06}m_e^{.62}m_p^{.82})$.  Barring migration, these are usually located at the ice line of the stellar system \cite{fernandes2019hints}, $a_\text{ice}\sim297.8\lambda^{43/30}M_{pl}^{2/3}/(\alpha^4m_e^{4/3}m_p^{1/3})$.  We can make an estimate for the number of gas giants per system as
\beq
N_\Jupiter\sim\frac{\int_{a_\text{ice}}^{a_\text{disk}}da 2\pi a \Sigma(a)}{M_\Jupiter}=3.70\times10^{-4}\,\frac{\alpha^{1.80}\,\beta^{.41}}{\kappa^{1/3}\,\lambda^{.47}\,\gamma^{.19}}\left(1-\frac{a_\text{ice}}{a_\text{disk}}\right)\theta\left(1-\frac{a_\text{ice}}{a_\text{disk}}\right)
\eeq
This is normalized to 2 for our values of the constants.  The factor $a_\text{ice}/a_\text{disk}=8.25\times10^7\kappa\lambda^{11/10}\alpha^{-4}\beta^{-4/3}\gamma^{1/3}$.  While this is small ($\sim.02$) for our values of the constants, the ice line exceeds the disk radius in about 10$\%$ of parameter space, depending on the habitability conditions used.

To determine the typical number of icy moons per gas giant, we'll first need the radius of gas giant planets.  From \cite{helled2023mass}, the density profile of a gas giant with a core of radius $R_\text{core}$ is given by $\rho(r)\sim \rho_\text{core}$ for $r<R_\text{core}$ and $\rho_0(1-(r/R_\text{core})^2)$ for $r>R_\text{core}$, with the value $\rho_0$ set to make the density continuous at $R_\text{core}$.  The core density $\rho_\text{core}\sim2\text{g/cm}^3$ is set by the degeneracy pressure of matter, and so scales as $\rho_\text{core}\sim \rho$, with $\rho\sim\alpha^3m_e^3m_p/8$ the density of matter.  Because of this, the radius can be approximated as $R_\Jupiter\sim(M_\Jupiter/\rho)^{1/3}\sim10.96\kappa^{1/9}\lambda^{.49}M_{pl}^{.94}/(\alpha^{1.60}m_e^{1.14}m_p^{.80})$.

Icy moons form around gas giants from a circumplanetary disk in much the same way that planets form from circumstellar disks.  In \cite{szulagyi2017effects}, it was found that circumplanetary disks have profile $\Sigma(a)\sim\xi_\Jupiter (2-p)M_\Jupiter/(2\pi r_\text{out}^{2-p}a^p)$, where $a$ is the distance from the planet and $\xi_\Jupiter\sim1.5\times10^{-3}$.  The parameter $p$ is taken to be $p=3/2$ for a ``minimal mass circumplanetary disk", though here, consistent with our treatment of circumstellar disks, we take $p=1$.  The disk inner edge is given by the Roche radius, $r_\text{in}\sim R_\Jupiter$, and the outer disk edge is set by the Hill radius, $r_\text{out}\sim .2 a_\Jupiter (M_\Jupiter/M_\star)^{1/3}$ \cite{ronnet2020formation}.  The typical moon mass is given by the isolation mass of this system, $M_\text{moon}\sim\xi_\Jupiter^{3/2}M_\Jupiter(a/r_\text{out})^{3-3p/2}$.  All the Galilean moons are within an order of magnitude of the isolation mass at their respective positions \cite{ronnet2020formation}. However, it should be noted that the formation process is rather complex, as this formation path can lead to resonant chains and instabilities, which can result in the coalescence of several moons into a larger one.  Indeed, with the exception of Titan, none of Saturn's moons seem to have formed in this way \cite{ronnet2020formation}.  Nevertheless, we use this formation pathway as a starting point in order to get a first estimate for the abundance of icy moons.

The resulting moon system will be close to maximally packed through this scenario, with each moon spaced several Hill radii apart.  While the total number of moons that result from this pathway is a rather complex (recursive) function of system parameters, a crude estimate can be obtained as the ratio of total disk size to the innermost Hill radius, $N_\text{moons/planet}\sim r_\text{out}/R_\text{Hill}(r_\text{in})$.  This neglects how the Hill radius changes as a function of radius.  This results in
\beq
N_\text{icy moons/gas giant}\sim\xi_\Jupiter^{-1/2}\left(\frac{r_\text{out}}{r_\text{in}}\right)^{2-p/2}=.072\,\frac{\lambda^{33/20}\,\gamma^{1/2}}{\alpha^{9/2}\,\beta^{1/2}}
\eeq
We normalize this to 3, after the bodies Europa, Ganymede, and Callisto for Jupiter, and Titan, Enceladus, and Dione for Saturn \cite{wikilakes}.  These are the largest but not the only known moons with subsurface oceans in the solar system, and there are most likely subsurface oceans on other moons that will be discovered in the future.

Since icy moon habitability is presumed to be contingent on a liquid subsurface ocean, we also delineate the conditions necessary for a subsurface ocean to persist.  We focus here on tidal friction as a source of heating, though in principle other sources, such as radioactivity, may contribute.  From \cite{tjoa2020subsurface}, the heat flux from tidal dissipation is given by
\beq
F_\text{tidal}=\frac{21}{8\pi}\frac{\epsilon_\text{tidal}\,e_\text{moon}^2\,G^{3/2}M_\Jupiter^{5/2}\,R_\text{moon}^3}{a_\text{moon}^{15/2}}
\eeq
Where $\epsilon_\text{tidal}\sim.005-.01$ is an efficiency factor.  The quantity $e_\text{moon}$ is the moon's eccentricity; because the eccentricity damping timescale of an isolated moon-planet system is so short, this mechanism is contingent on the presence of multiple moons within the system \cite{heller2014formation}.

The thickness of the icy crust from this heat source is given by $D_\text{ice}\sim k \Delta T/F_\text{tidal}$, where $k$ is the thermal conductivity.  We take $\Delta T\sim E_\text{vib}$ to be the melting point of water, neglecting the residual surface temperature and and increased insulation effect from a regolith layer, and ignoring any additional heat sources, such as flux from the parent star \cite{tjoa2020subsurface}.  Here, we approximate the thermal conductivity as $k\sim0.01\alpha^3m_e^{11/4}/m_p^{3/4}$; though a more thorough treatment would take into account the temperature dependence (as in \cite{andersson2005thermal}), this would not alter our results appreciably.  The ice thickness is then given by
\beq
D_\text{ice}\sim4.09\times10^{14}\,\frac{\lambda^{.18}}{\kappa^{1/3}\,\alpha^{2.20}\,m_e^{.34}\,m_p^{.34}}
\eeq
This is normalized to 20 km for Europa.  This ratio then becomes
\beq
\frac{D_\text{ice}}{R_\text{moon}}=2.86\times10^{15}\,\frac{\lambda^{.24}\,\beta^{.62}\,\gamma^{1.42}}{\kappa^{4/9}\,\alpha^{2.11}}
\eeq

The condition for a subsurface to exist is that not all the water on the moon be frozen.  The ocean depth of icy moons is uncertain, but can be estimated to be about $\sim5$ times the thickness of the icy crust\cite{howell2020nasa}.  The number of icy moons per star is then
\beq
N_\text{icy moons/star}=2.7\times10^{-5}\,\frac{\lambda^{1.18}\,\gamma^{.31}}{\kappa^{1/3}\,\alpha^{2.70}\,\beta^{.09}}\left(1-\frac{a_\text{ice}}{a_\text{disk}}\right)
\theta\left(1-\frac{a_\text{ice}}{a_\text{disk}}\right)
\theta\left(.4-\frac{D_\text{ice}}{R_\text{moon}}\right)
\eeq


Restricting attention to our universe, the bound on the the relative habitability of icy moons is
\beq
r^{h|u}_\text{icy moons}<8.1
\eeq
Whereas when considering throughout the multiverse, we have
\beq
r^h_\text{icy moons}<19.3\quad r^{h\cap\,\mU}_\text{icy moons}<1.2
\eeq
From this, we see that our universe is actually pretty abundant in icy moons compared to the average throughout the multiverse. However, when considering the probability that we are both on a temperate and terrestrial planet instead of an icy moon and in a universe like ours, the bounds on the relative habitability of icy moons are stronger than when considering their abundance in our universe in isolation.

\section{Rogue Planets}\label{rogue}

We now turn our attention to rogue planets, which are planets that are ejected from their host star system during the process of planet formation.  While traditionally expected to not have properties conducive to life, it was pointed out in \cite{stevenson1999life} that because of their low temperatures they may retain large hydrogen and helium atmospheres, potentially providing surface temperatures capable of hosting liquid water for geologically long time periods.  Additionally, it was found in \cite{abbot2011steppenwolf} that rogue planets may form thick ice coverings that maintain subsurface liquid oceans for billions of years.  Lastly, \cite{debes2007survival} found that when a planet is ejected, any moons it may possess are (slightly) more likely than not to remain bound to the planet, with in fact highly increased eccentricity, providing an additional source of tidal heating that may help maintain liquid water.

Rogue planets are expected to be prevalent in our galaxy, and indeed have been observed \cite{microlensing2011unbound}, but estimates of their abundance are currently extremely broad: a recent estimate places the number of rogue planets per star as $21^{+23}_{-13}$ \cite{sumi2023free}.  If the number is toward the higher estimate, we then have reason to believe on the basis of abundances alone that rogue planets are less habitable than bound planets.  Within the multiverse context, we ask two questions: How do the numbers of bound and rogue planets depend on constants? Are both generic features of planetary system evolution, or do we inhabit a universe that anomalously produces one or the other?

Planet formation is a complex, chaotic process, and a fully analytical treatment of the distribution of rogue planets ejected during the formation process is so far lacking.  Nevertheless, we may incorporate some key facts known about the formation of rogue planets, garnered from observations, theory, and simulations, to arrive at a crude estimate for the average number of rogue planets per star as a function of system properties and fundamental constants.

In \cite{matsumoto2020ejection}, it was found that planets are ejected only when the ejection mass is exceeded, which is defined by the condition $GM_\text{ej}/R=GM_\star/a$, or $M_\text{ej}\sim M_\star^{3/2}/(\rho^{1/3}\,a^{3/2})$.  We can use this to define the ejection radius $a_\text{ej}$, which corresponds to the distance from the star at which the oligarchic mass is greater than the ejection mass, so that ejections occur.  In \cite{ma2016free} it was found that most planet ejections do indeed occur in the outer parts of planetary systems, lending credence to this concept, although ejections certainly do occur inward of this point.  Equating the ejection and oligarchic masses gives
\beq
a_\text{ej}=.14\,\frac{\lambda^{1/3}\,M_\text{pl}}{\kappa^{2/5}\,\alpha^{3/5}\,m_e^{3/5}\,m_p^{7/5}}
\eeq
which for our constants and for solar mass stars is 1.84 AU.  We also note that in universes where this quantity is larger than the disk radius, ejection plays a minimal role, and rogue planets are rare.

The final surface mass density of planets as an outcome of the planet formation process can then be found as $\Sigma_\text{planet}(a)\sim \text{min}(M_\text{olig},M_\text{ej})/(2\pi R_\text{Hill} a)$.  The number of ejected planets can then be approximated as 
\beq
N_\text{rogue}\sim \int_{a_\text{ej}}^{a_\text{disk}}da\,2\pi a\frac{\Sigma(a)-\Sigma_\text{planet}(a)}{M_\text{iso}(a)}
\eeq
We have made the assumption that all of the gas/dust ultimately is incorporated into isolation mass planets before they are ejected, and that this quantity can be treated continuously.  More importantly, we have made the simplification that all ejected planets are ejected before they grow much larger than their isolation mass.  In truth, planets could be ejected at any point in the formation process, but simulations find that the masses of ejected planets are very close to the initial protoplanet mass \cite{matsumoto2020ejection}.  At any rate, the distribution of rogue planet masses is expected to be much more skewed toward lower masses than the population of bound planets, due to the asymmetry of the ejection process \cite{gould2022free}.

The above integral can then be evaluated as
\beq
N_\text{rogue} = \left[.1\left(\frac{\alpha\,\beta}{\kappa}\right)^{3/10}-3.4\times10^{-5}\left(\frac{\alpha\,\beta}{\kappa}\right)^{1/2}\right]\theta(a_\text{disk}-a_\text{ej})
\eeq
Two things stand out in this expression: the first is that $N_\text{rogue}$ only depends on the particular combination of parameters $\alpha\beta/\kappa$, a consequence of the fact that the integrated expression only depends on the quantity $a_\text{ej}/a_\text{disk}\sim4\times10^4(\kappa/(\alpha\beta))^{3/5}$.  The second is that the number of ejected planets does not depend on stellar mass.  The fraction of ejected material is also found to be independent of stellar mass; this is in conflict with \cite{ma2016free}, which finds this to be a decreasing function of stellar mass.  However, this independence is a consequence of our ejection radius formalism, and is quite independent of other assumptions, for instance if we instead take $\Sigma(a)\propto a^{-p}$.  It arises because $a_\text{ej}$ is a function of only $M_\star$, $\rho$, and $a_\text{disc}(M_\star,\rho_\text{gal})$.  Then, dimensionally, $a_\text{ej}\sim M_\star/(\rho^q\rho_\text{gal}^{1/3-q})$ for some $q$, and $a_\text{ej}/a_\text{disk}\sim(\rho_\text{gal}/\rho)^q$.  The result then follows from the fact that $N_\text{ej}$ is a function solely of this last ratio of length scales.

Taken all together, the relative habitability of rogue planets when restricting to our universe is bounded as
\beq
r^{h|u}_\text{rogue}<.31
\eeq
whereas when considering their abundance throughout the multiverse, the bounds are much stronger:
\beq
r^h_\text{rogue}<5.1\times10^{-4}\quad r^{h\cap\,\mU}_\text{rogue}<1.0\times10^{-8}
\eeq
This much stronger bound reflects that rogue planets are much more pervasive than bound planets throughout the multiverse, as compared to our universe.  This is driven by the fact that the habitability of rogue planets cannot be reliant on photosynthesis, as the prospects for this to develop on these systems are quite poor \cite{moonlight}.  As such, this opens up a much wider range of parameter space where the habitability of rogue planets would be unaffected, compared to the habitability of bound planets, which is taken to be reliant on photosynthesis via the {\bf yellow} habitability condition.  We note that we have neglected the possibility raised in \cite{ACBplanets} that occurs when galactic density is much greater; in universes like these, incident starlight flux is equal to the amount Earth receives, rendering essentially every planet, both rogue and bound but beyond the traditional circumstellar habitable zone, capable of retaining surface liquid water.

\section{Water}\label{water}

\subsection{Properties of water}

The properties of water have long been recognized as uniquely suited for life as we know it on Earth, and were even the subject of some of the earliest anthropic arguments.  The 1913 book \emph{The Fitness of the Environment} written by Lawrence Henderson \cite{henderson1913fitness} pointed out a number of features of water that were deemed crucial for life: 1) its high thermal capacity, which helps regulate seasonal and daily temperature changes on Earth, 2) its high melting point compared to other liquids, allowing it to remain liquid on Earth's surface, 3) its high latent heat of evaporation, which moderates Earth's temperature, 4) the nearly unique fact that ice floats, preventing bodies of water from completely freezing through, 5) its property as a universal solvent, allowing for complex electrochemistry, 6) its high surface tension, which concentrates reactants into smaller droplets and facilitates capillary action, and 7) the fact that carbon dioxide is capable of existing both in soluble and gaseous form, facilitating the transport of this molecule for biological needs and regulating CO$_2$ levels in the atmosphere (if carbonate were not deposited over geologic time, the world's limestone cliffs would still be in the air, as on Venus).

Work over the subsequent century has largely corroborated this view that the properties of water are somewhat exceptional compared with other liquids; a notable work along these lines is the conference proceedings published in \cite{lynden2010water}.  However, the beneficence of some of these properties has been called into question, and it has even been suggested that some of these properties make water an adversarial environment for life to take hold \cite{pace2010fine}.  In \cite{benner2010constraining} it was pointed out that water's high melting point restricts the ``habitable zone'' around a star to a narrow band, and if the melting point were lower, a greater fraction of rocky bodies within the universe would contain liquid water.  While being a universal solvent may facilitate reactions, it also necessitates constant repair, as many biomolecules, including DNA and proteins, readily disintegrate in water \cite{benner2014paradoxes}.  Sea ice raises the albedo of Earth appreciably, leading to a positive climate feedback that exacerbates temperature fluctuations and has triggered worldwide glaciations in the past \cite{pierrehumbert2011climate}.  Carbon dissolvability may be pathological too, as the bicarbonate form it takes when dissolved is unreactive \cite{benner2010constraining}.  In lieu of this, it is safe to say that the overall impact on habitability of these aspects of water are currently under debate- it may well be that life makes use of water on Earth as the most abundant liquid, despite its flaws.

Furthermore, as tabulated in Chapter 7 of \cite{schulze2004life}, many other liquids, themselves abundant throughout the universe, possess some of the same properties as water.  In Table \ref{liquid_table} we reproduce some of these properties for some of the most abundant liquids for convenience.  Which among these liquids are capable of supporting life? This depends on which of the properties of water are essential.  If these liquids turn out to be suitable for life, it may well be that the majority of life within the universe takes place in a solvent other than liquid water \cite{ballesteros2019diving}.

\begin{table*}
	\centering
        \begin{tabular}{|c|c|c|c|c|c|c|c|c|c|}
	\hline
	Liquid & Formula & $T_m$ & $\epsilon$ & $\nu$ & $D$ & $\sigma$  & ice floats & $R$ & location\\
	\hline
        
        water & H$_2$O &  0 & 80 & 9.6 & 1.9 & 72 & yes & 1 & Earth \\
        ammonia & NH$_3$ & \cellcolor{blue!10}-78 & \cellcolor{red!15}17 & \cellcolor{green!15}2.7 & 1.5 & \cellcolor{red!15}20 & no & 0.25 & Jupiter \\
        hydrogen cyanide & HCN & -13 & \cellcolor{green!15}115 & \cellcolor{green!15}2 & \cellcolor{green!15}3 & \cellcolor{red!15}18 & no & 0.14 & Jupiter, Titan \\
        hydrogen sulfide & H$_2$S & \cellcolor{blue!10}-86 & \cellcolor{red!15}6 & \cellcolor{green!15}4.3 & \cellcolor{green!15}10 &  & no & 1.31 & Io \\
        sulfuric acid & H$_2$SO$_4$ & 10 & \cellcolor{green!15}101 & \cellcolor{red!15}260 & \cellcolor{green!15}2.7 &  & no & 0.42 & Venus$^\dagger$ \\
        methane & CH$_4$ & \cellcolor{blue!10}-182 & \cellcolor{red!15}2 & \cellcolor{green!15}0.01 & \cellcolor{red!15}0 &  & no$^*$ & 0.62 & Titan \\
        ethane & C$_2$H$_6$ & \cellcolor{blue!10}-172 & \cellcolor{red!15}2 & \cellcolor{green!15}0.01 & \cellcolor{red!15}0 &  & no$^*$ & 1.25 & Titan \\
        nitrogen & N$_2$ & \cellcolor{blue!10}-210 & \cellcolor{red!15}1 & \cellcolor{green!15}2 & \cellcolor{red!15}0 & \cellcolor{red!15}11 & no & 1.96 & Triton \\
        \hline
	\end{tabular}
        \caption{Properties of the most abundant liquids in our solar system.  Reproduced from \cite{schulze2004life}.  Cells colored green may be considered as more beneficial to life than water, red less, and blue uncertain.
 $T_m$: melting temperature ($^\degree$C). $\epsilon$: dielectric constant. $\nu$: viscosity ($10^{-3}$P). $D$: dipole moment (debye). $\sigma$: surface tension ($10^{-3}J/m^2$). $R$: relative abundance.\\ $^\dagger$: present in aerosols \cite{gao2014bimodal}\\ $^*$: porous ice may float \cite{yu2024fate}}
	\label{liquid_table}
\end{table*}

Today, many of water's special properties are known to stem from the fact that it is a ``slightly frustrated tetrahedral liquid'' \cite{finney2004water} - stereochemically, water molecules behave as if they were tetrahedra (four sided pyramids, each side a nearly equilateral triangle), with two of the vertices representing the excess electrons of the oxygen atom, and the other two vertices representing the positively charged hydrogen atoms.  The positively charged vertices of one water molecule interact readily, though weakly, with the negative vertices of other water molecules nearby, resulting in a highly dynamic supramolecular structure \cite{kholmanskiy2019supramolecular}.  The fact that the bond angle between water's two OH bonds is close, but not quite equal to, the angle of a regular tetrahedron (104.5$^\degree$ versus 109.5$^\degree$) poises the system close to a critical point, with long range correlations that exist on the threshold between a regular packing and fluid state \cite{finney2004water}.

This tetrahedral nature of water molecules results from $sp3$ bond hybridization, and results in an H bond that is much weaker than ordinary covalent or ionic bonds \cite{franks2010water}.  Its effects are quite routinely captured by the effective Stillinger-Weber potential, where the bond strength is input as $E_H=23$ kJ/mol (.24 eV) \cite{russo2018water}. In \cite{chaplin2010water}, many of the unique properties of water were traced to this particular value of the bond strength.  Therein, they used a combination of methods (altering conditions, theoretical modeling, using different isotopes, and measuring different liquids) to estimate thresholds beyond which these properties no longer hold.  In Table \ref{eh_table} we reproduce a simplified version of their findings.  Additionally, they find that the low viscosity of water is highly sensitive to changes in the bond strength, which may have consequences for the rapidity of reactions in solution.

\begin{table*}
	\centering
        \begin{tabular}{|c|c|}
	\hline
	$\Delta E_H$ & Effect \\
	\hline
	$-11\%$ &  K$^+$ becomes kosmotropic\\
	$-2\%$ &  ice sinks\\
	$+3\%$ &  viscosity increases $23\%$, diffusivity reduced $19\%$\\
	$+11\%$ &  Na$^+$ becomes chaotropic\\
        \hline
	\end{tabular}
        \caption{Reproduced from \cite{chaplin2010water}. The effects of altering the bond strength of water at $37^\degree$C and 1 atm pressure.}
	\label{eh_table}
\end{table*}

In the multiverse framework, each of these thresholds can be incorporated into the calculation of habitability as a function of the fundamental constants, under varying assumptions about their importance.  Comparing the resulting probabilities of being in our universe gains us predictions about which of these properties of water are important for life, and consequently which other liquids are capable of supporting life.  For this, we make use of the formula $E_H\sim\alpha^2\,m_e$ \cite{PL}.

The effects of each of these conditions are reported in Table \ref{water_bayes_table}.  The \emph{effect} column outlines the macroscopic property considered, with the first row denoting the stance that all water's special properties are inconsequential. The \emph{condition} column indicates the bounds on the relevant physical quantity to allow that condition.  The $\mathcal B$ column gives the relative Bayes factors when incorporating that condition into the multiverse habitability calculations, defined as $\mathcal B= P(\mU,r$=$1)/P(\mU|h)$. The first Bayes factor reported restricts the other habitability conditions to the preferred default {\bf yellow + C/O + S}. The second Bayes factor reported allows complementary habitability conditions to vary.  The conditions that yield the maximum Bayes factor are reported in the \emph{best combination} column, with the following abbreviations: {\bf bio}: habitability is restricted to systems with stellar lifetime longer than the biological speciation timescale \cite{mc1}, {\bf area}: habitability is proportional to planet area \cite{mc2}, {\bf terr}: habitability is restricted to planets that can retain heavy but not light gases \cite{mc2}, {\bf temp}: habitability is restricted to planets capable of hosting surface liquids \cite{mc2}. The last column reports the bounds on relative habitability that can be inferred from these analyses.  The within-universe bounds are computed from the relative abundance column in Table \ref{liquid_table}, and are conservative in that it only compares the liquids not possessing the property under consideration to water, rather than all other liquids that possess those properties.  For the multiverse bounds, we have restricted to the {\bf yellow + C/O + S} conditions.  For many of the habitability conditions, no bounds can be found on $r^{h\cap\,\mU}$, as the baseline probability when assuming these conditions is already less than $p_c$, and thus no value of relative habitability would make these conditions consistent with our observations.

From this, we can draw a number of implications for the properties of water which are expected to have a strong bearing on its habitability.  We see from the second row that, when considering the {\bf yellow + S + C/O} condition, the {\bf solvent} condition drops the Bayes factor considerably, indicating that taking this aspect of water to be important is incompatible with our observations within the multiverse context.  However, this restriction is relaxed if other habitability conditions turn out to be more appropriate, so that the solvent property may indeed be essential to life if habitability is proportional to stellar lifetime and planet area, rather than entropy.  We also find the bounds on the relative habitability of liquids that are not universal solvents to be stronger in the multiverse setting than the within universe context, but still compatible with being more habitable than water. Thus we make no strong predictions about the habitability of nonsolvent liquids, such as methane, ethane and nitrogen.

Additionally, we find the {\bf ice floats} condition to universally give worse Bayes factors, indicating that this feature is not expected to be necessary for habitability, independent of additional habitability assumptions.  From this, we do not expect life to be precluded in liquids whose ice sinks, which is all other liquids aside from water (with the possible exception of methane/ethane, as evidenced by the floating islands on Titan \cite{yu2024fate}).  These liquids may of course be unsuitable for life for other reasons.

The {\bf viscosity} condition leads to a small drop in Bayes factors, and very weak bounds on relative habitability.  On this basis, sulfuric acid, which has favorable properties except for high viscosity, can be expected to be suitable for biochemistry, again in the absence of some other precluding factor.

Lastly, the {\bf ice+viscosity} condition, which narrows to the most restrictive habitable range we consider, yields worse Bayes factors, unless habitability is taken to be proportional to planetary surface area rather than entropy received via starlight.  Generically, we see that the two sided bounds, where $E_H$ is bounded both from above and below, place the most stringent bounds on relative habitability, while lower bounds on $E_H$ lead to the lowest probabilities of our observations.

\begin{table*}
	\centering
        \begin{tabular}{|c|c|c|c|c|}
        \hline
        effect & condition & $\mathcal{B}_\text{ref}$/$\mathcal{B}_\text{max}$  & best combination & $r^{h|u}/r^h/r^{h\cap\,\mU}$\\
	\hline
	null & - & 1 / 1 & {\bf photo bio S C/O} & -\\
	solvent & $.89<E_H<1.11$ & .045 / .67 & {\bf bio area} & 5.0/1.6/-\\
	ice floats & $.98<E_H$ & .0095/.037 &  {\bf terr area C/O} & 4.7/6.8/-\\
	viscosity & $E_H<1.03$ & .17/.41 &  {\bf temp area C/O} & 45.2/58.3/-\\
	ice + viscosity & $.98<E_H<1.03$ & .066/.83 & {\bf temp area C/O} & 4.7/.37/-\\
	abundance & $f_\text{temp}$ & .80/.94 & {\bf yellow area C/O} & -\\
	Al-26 & $.1<\hat t_{\text{Al-26}}<10$ & 1.3/2.1 & {\bf terr area C/O} & .27/.082/.014\\

        \hline
	\end{tabular}
        \caption{Thresholds for different changes to the properties of water. Columns explained in text.}
	\label{water_bayes_table}
\end{table*}

\subsection{Melting Temperature}

The high melting point of water is beneficial for the retention of liquid surface water in the \emph{currently defined} habitable zone.  However, as pointed out in \cite{benner2010constraining}, if the melting temperature of water were lower, a much larger fraction of rocky bodies would have conditions required to retain liquid on their surface.  Here we take this factor into account in our multiverse calculations.  Here, we only consider the potential for liquid water on a surface, not the water abundance that results from planet formation scenarios.  For more details on how this may depend on fundamental constants as a function of water origin scenario, see \cite{mc6}.

To proceed, we need the fraction of rocky bodies orbiting a star of size $\lambda$ that possess a surface temperature $T$.  For this, we use the surface abundance distribution as a function of semi-major axis $a$, $\Sigma(a)\propto1/a$, as discussed in section \ref{icy}.  As a first approximation, we are making the assumption that the size distribution of bodies does not appreciably depend on semimajor axis, though of course it does.  

Since planets orbit in a disk, the above density may be coupled with the Jacobian factor $2\pi a$ to result in a probability distribution of the occurrence of rocky bodies as a function of semimajor axis $p(a)\sim $ const, and a cumulative distribution function $c(a)=a/a_\text{disk}$.  Here, the outer limit of the semimajor axis distribution is set by the galactic stellar density, $a_\text{disk}\sim n_\text{gal}^{-1/3}$, and we can relate the semimajor axis to surface temperature via $a\sim L_\star^{1/2}/T^2$, since the temperature of a body is dictated by radiative equilibrium with the host star.  We can then find the fraction of rocky bodies between $T_\text{freeze}=\eta_-T_\text{mol}$ and $T_\text{boil}=\eta_+T_\text{mol}$ as
\beq
f_\text{temp}\sim\left(\frac{1}{\eta_-^2}-\frac{1}{\eta_+^2}\right)\frac{n_\text{gal}\,L_\star^{1/2}}{T_\text{mol}^2}\sim\frac{\kappa\,\lambda^{17/12}\,\gamma^{1/2}}{\alpha^5\,\beta^2}
\eeq
Both $T_\text{freeze}$ and $T_\text{boil}$ are expected to scale linearly with $T_\text{mol}$, making $\eta_\pm$ constants.  This is corroborated by the observation in \cite{chaplin2010water} that the melting and boiling points of water scale linearly with hydrogen bond strength, where they noted that at $0^\degree$C, water would boil if $E_H$ were decreased by $29\%$ and freeze if increased by $18\%$.

The salient feature of this distribution is that $f_\text{temp}\propto1/T_\text{mol}^2$, so that as the melting temperature decreases, the fraction of bodies capable of retaining liquid surface water increases drastically.

The results of taking the habitability of a universe to be proportional to this factor are given in Table \ref{water_bayes_table}.  The resulting Bayes factor is .80 when restricting to the {\bf yellow + S + C/O} habitability condition, and .94 when allowed to scan over additional conditions.  From this, we conclude that this weighting is compatible with the multiverse scenario, but is neither favored nor disfavored.

\subsection{Aluminum-26 Heating and Waterworlds}

The above discussion was focused on the properties of water, not necessarily its availability on inner rocky planets.  Indeed, this may be an important factor to consider; here, we point out one factor that may or may not be of crucial importance in dictating the water abundance of inner planets- the decay of aluminum-26.

Aluminum-26 is a radioactive isotope that is produced in supernova and the winds of heavy stars, with a half-life of 717,000 years.  This being about the timescale of planet formation, the decay of this isotope is the dominant radioactive energy source during the early stages of planet formation \cite{grimm1993heliocentric}.  This decay plays an important role in the evolution of planetesimals; it acts as an energy source that causes any water present in a planetesimal to evaporate, dehydrating the material \cite{lichtenberg2019water}.  This results in a drastically lower water content of the planetary building blocks, and is argued to be necessary for the formation of rocky rather than water-rich worlds.  In \cite{lichtenberg2019water} it was suggested that only inner planets in regions of the galaxy with active supernovae during the planet formation phase are expected of possess rocky planets, rather than waterworlds.  In \cite{gounelle2015abundance}, it was estimated that only about $1.4\%$ of star systems are as rich in aluminum-26 as ours.

From the multiverse point of view, this claim is interesting because the half-life of aluminum-26 is quite sensitive to variations in the fundamental constants, as first discussed in \cite{al26}.  There, the following expression for the half life was found to be (Taylor expanded around the observed values; the full expression is contained therein):
\beq
\hat t_{\text{Al-26}} = 1+7.29\left(\hat\beta-1\right)+26.03\left(1.92\hat\delta_d-.92\hat\delta_u-1\right)-75.32\left(\hat\alpha-1\right)
\eeq
In this expression we use hats to denote the values of the constants normalized to our observed value, $\hat x = x/x_\text{obs}$, where $x_\text{obs}$ is the observed value.  With this, we can compare the two different hypotheses, of whether or not the decay of aluminum-26 is relevant in setting the water content of rocky planets, by including a condition that the half life still remain within an order of magnitude of the planet formation timescale, $\mathbb H_\text{Al-26} = \theta(\hat t_\text{Al-26}-.1)\theta(10-\hat t_\text{Al-26})$.  If the decay were to happen too quickly, the isotopes would have released their energy during transit from their initial source, whereas if the decay were too slow, the energy would be released after the planet formation process was complete. This simple condition neglects any variation in the planet formation timescale which may concurrently occur, but serves as a useful starting point to investigate the importance of this effect in a multiverse context.

With this habitability condition, we find that the relative Bayes factor when restricting to the {\bf yellow + C/O + S} condition is 1.3. The maximum Bayes factor when scanning over different habitability conditions occurs for the {\bf TERR + area + C/O + al26} conditions, and is 2.1.  Thus, the idea that aluminum-26 decay is important for habitability is compatible with, and even mildly favored by, the multiverse hypothesis.  

To the extent that this mechanism is operational in determining the water abundance of interior planets, this also allows us to test the expected relative habitability of waterworlds.  From the abundance estimates on these systems quoted above, we can infer within our universe that 
\beq
r^{h|u}_\text{waterworld}<.27
\eeq
Whereas when considering the full multiverse context, we find
\beq
r^h_\text{waterworld}<.082,\quad r^{h\cap\mU}_\text{waterworld}<.016
\eeq
Since waterworlds are the default outcome of planet formation in this setup, we find that the bounds on their relative habitability are stronger in the multiverse context than compared to what we find in our universe alone.

\section{Discussion}

Our main results from each section are summarized in Table \ref{r_table}, where we show that the expectation for the relative habitability for various exotic environments, given the fact that we find ourselves to be humanlike, can be greatly altered when considering a multiverse context, rather than only restricting our attention to our universe.  Generally, this is purely a statistical statement, which depends only on the relative abundance of the two types of locations, and so the bounds on relative habitability are greatest when the exotic location is significantly more abundant in the multiverse setting than within our universe.

\begin{table*}
	\centering
        \begin{tabular}{|c|c|c|c|}
        \hline
        exotic environment & $r^{h|u}$ & $r^h$ & $r^{h\cap\mU}$ \\
	\hline
        condition & $P(h\,|\, u)>p_c$ & 
        $P(h)>p_c$ & 
        $P(h\cap\mathcal U)>p_c$\\
	\hline
	tidally locked ({\bf yellow})& 507.4 & 2.0 & .38\\
	tidally locked ({\bf photo})& .91 & .68 & .13\\
        binary stars & 19.0 & 25.9 & 5.0\\
	icy moons & 8.1 & 19.3 & 1.2\\
	rogue planets & .31 & $5.1\times10^{-4}$ & $1.0\times10^{-8}$\\
	nonsolvents & 5.0 & 1.6 & -\\
	ice-sinking fluids & 4.7 & 6.8 & -\\
	viscous fluids & 45.2 & 58.3 & -\\
	viscous, ice-sinking fluids & 4.7 & .31 & -\\
	waterworlds & .27 & .082 & .014\\
        \hline
	\end{tabular}
        \caption{Bounds on the relative habitability of various systems by considering the relative abundances of these systems, and the probability of being humanlike within our universe, the probability of being humanlike in the multiverse, and the probability of being humanlike and within a universe like ours.  All numbers in this table use threshold probability $p_c=.05$.}
	\label{r_table}
\end{table*}

Because these results rely on several assumptions, we reiterate them here.  Firstly, we have assumed that the relative habitability factor $r$ for each exotic environment is independent of physical parameters.  This is a natural assumption, in the absence of a physical argument otherwise, and was necessary for the tractability of the calculations, but could very well be untrue.  For example, though binary systems may very well be just as habitable as lone stars in our universe, perhaps in universes with different constants they are more prone to instability, rendering them uninhabitable, and thus weakening our bounds.  These types of considerations can always be folded into our calculations, at the expense of increased computational complexity.

Secondly, we have only varied the microphysical constants, and have held the cosmological constants fixed.  We are extending our work to incorporate varying cosmological parameters in a future work.

Lastly, the bounds we find are dependent on which auxiliary habitability conditions we assume.  For all but the second row of this table, the baseline {\bf yellow + C/O + S} condition was used, where habitable systems are restricted to yellow stars with C/O$<1$, and where habitability is proportional to total entropy received.  Generically, resulting probabilities can be highly sensitive to the habitability assumptions that are made, and so if these turn out to be disfavored, the bounds for each of these exotic environments must be recomputed.

This work addresses one of the main charges of the multiverse hypothesis, that it ordinarily employs quite strictly humanlike assumptions about the nature of habitability \cite{schellekens}.  In general, since observations in the multiverse can only be predicted in a statistical manner, we may only test the multiverse by determining if our observations are reasonably typical or not.  Most of the work on this matter has been framed around the observation of our particular set of physical constants.  However, the salient features of our local environment is another important class of observations to consider, and indeed we find that these have the potential to be more constraining than considering only physical constants alone.

The result of this work is a list of predictions for how habitable various exotic environments can be.  In particular, the bounds on the relative habitability of rogue planets and waterworlds are much stronger in the multiverse context than within our universe, as these systems are the majority of sites throughout the multiverse.  These exotic systems considered have a range of relative habitability values between $r^{h\cap\mU}$ and $r^{h|u}$ that would make our presence as humanlike typical within our universe but atypical within the multiverse, if the relative habitability turns out to lie within this range.  Additionally, we found that taking either the fact that ice floats or that water is a universal solvent as essential for life was incompatible with our observations (the former, independently of other habitability conditions assumed, the latter, with some caveats).  The recommendation to test the multiverse hypothesis, then, is to go out and check the habitability of each of these environments.  If any of them do turn out to be more habitable than the multiverse predicts, this would falsify the multiverse hypothesis, in some case to very high statistical precision.


\smallskip

\bibliographystyle{apalike}
\bibliography{yellowbib}

\end{document}